\documentclass[twocolumn,aps,pre,preprintnumbers,amsmath,amssymb,nofootinbib]{revtex4-1}

\usepackage{graphicx}% Include figure files
\usepackage{dcolumn}% Align table columns on decimal point
\usepackage{bm}% bold math

\begin{document}

\title{Charged Bilayer Membranes in Asymmetric Ionic Solutions: \\ Phase Diagrams and Critical Behavior}% Force line breaks with \\

\author{Naofumi Shimokawa\footnote{e-mail:
\texttt{nshimo@iis.u-tokyo.ac.jp}}}
\affiliation{%
Institute of Industrial Science, University of Tokyo,
Tokyo 153-8505, Japan
}%
\author{Shigeyuki Komura}
\affiliation{%
Department of Chemistry,
Graduate School of Science and Engineering,
Tokyo Metropolitan University, Tokyo 192-0397, Japan,
and Kavli Institute for Theoretical Physics China, CAS, Beijing 100190, China
}%
\author{David Andelman}
\affiliation{%
Raymond and Beverly Sackler School of Physics and Astronomy,
Tel Aviv University, Ramat Aviv 69978, Tel Aviv, Israel,
and Kavli Institute for Theoretical Physics China, CAS, Beijing 100190, China}%

%\date{July 12, 2011 --- revised after refs. reports}

\begin{abstract}
We consider the phase separation in an asymmetrically charged lipid bilayer
membrane consisting of neutral and negatively charged lipids that are in
contact with in and out ionic solutions having different ionic strengths.
The two asymmetric leaflets are coupled through  electrostatic
interactions.
Based on a free-energy approach, the critical point
and phase diagrams are calculated for different ionic strengths of
the two solutions and  coupling parameter.
An increase of the coupling constant
or asymmetry in the salt concentration between the in and out solutions yields
a higher phase separation temperature because of  electrostatic interactions.
As a consequence, the phase-coexistence region increases for strong screening (small Debye length).
Finally, possible three-phase coexistence regions in the phase diagram are
predicted.
\end{abstract}

%\keywords{Suggested keywords}%Use show keys class option if keyword
                              %display desired
\maketitle
%\baselineskip=24pt

%%%%%%%%%%%%%%%%%%
%  Introduction  %
%%%%%%%%%%%%%%%%%%
\section{Introduction}
\label{intro}

Model membranes consisting of mixtures of saturated lipids, unsaturated
lipids, and cholesterol have received considerable attention in recent years
because they can be regarded as
model systems of biological cell membranes.
Below a certain temperature, the membrane undergoes a lateral phase separation
between a liquid-ordered phase (rich in saturated lipid and cholesterol) and
a liquid-disordered phase (rich in unsaturated lipid).
The resulting lipid domains are sometimes called ``rafts", and are believed to play an
important role in various bio-cellular processes such as signal transduction
and cooperative membrane trafficking~\cite{SI}.

A large number of experimental studies have been
carried out to further explore the consequences of domain formation
in model membranes and their relation with biomembrane functioning.
They include, among others, studies of domain morphology~\cite{Keller1,Keller2},
domain budding~\cite{Webb}, growth dynamics of domains~\cite{Keller3}, and
formation of periodic lateral structures~\cite{Rozovsky}.
In particular, we note that lateral
phase segregation was directly observed for lipids at the air-water interface and for giant
lipid vesicles using fluorescence
microscopy~\cite{Keller1,Keller2,Webb,Keller3,Rozovsky}.
Furthermore, in an attempt to understand these experimental findings,
several theoretical models have been proposed~\cite{LD,KSA,PS1}.

More recently, several experimental works have been conducted in order to
understand how electrostatic interactions affect the phase separation of
model membranes composed of charged and neutral lipids~\cite{SHSY,Dimova}.
As charged lipids are everpresent in biomembranes, the role of electrostatic
interactions is important also for biological cells.
The phase-coexistence region was reported to be
fully suppressed in lipid bilayers consisting of a three-component mixture of neutral
saturated lipid DPPC (1,2-dipalmitoyl-sn-glycero-3-phosphocholine),
negatively charged  unsaturated lipid DOPS (1,2-dioleoyl-sn-glycero-3-phospho-L-serine),
and cholesterol, because of the electrostatic repulsion between
charged lipids~\cite{SHSY}. Furthermore, the extent of the phase-coexistence region
appears when salt (CaCl$_2$) is added, an effect that
can be understood by the screening of electrostatic interactions in presence of added salt.
In a related experimental study, it was shown that
for mixtures of DPPC, negatively charged DOPG (L-$\alpha$-1,2-dioleoyl-sn-glycero-3-phosphoglycerol)
and cholesterol,
the phase-separation temperature becomes higher as salt is added~\cite{Dimova}.
Such effects were also considered in several theoretical studies~\cite{MHB,HMB,MBM}.

Another important aspect of biomembranes is the compositional asymmetry
between their inner and outer leaflets.
Collins and Keller investigated the phase behavior in asymmetric lipid
bilayers, and showed that the domain formation in each leaflet is
coupled through an inter-leaflet interaction~\cite{CK}.
In some cases, it was reported that phase separation in one of the
leaflets induces a phase separation
in the second leaflet, while in other cases, the lack of phase separation
in one leaflet suppresses domain formation in the second leaflet.
Theoretical models based on regular solution theory~\cite{WLM} or
Landau theory~\cite{PS2} attempted to take into account
inter-leaflet interactions and explored their consequences on the bilayer phase diagram.

In related works based on the Poisson-Boltzmann theory,
a  model describing the immiscibility transition in {\it asymmetric and
charged} membranes was proposed by May and coworkers~\cite{BM,WM}.
Their model took into account
an inter-leaflet {\it electrostatic} coupling between two rigid and charged planes modeling a bilayer membrane.
The spinodal line, characterizing the phase coexistence, and the critical
point were derived and  depend on the ionic strength of the solution as well as
on the inter-leaflet electrostatic coupling.
One of the main conclusions was that a stronger inter-leaflet coupling enhances
the lateral phase separation.
However, it was equally assumed by the authors that the two aqueous solutions in contact with the two
leaflets have the same salt concentration. We note that this assumption is an over-simplification for
cellular biomembranes, where due to ionic channels and other active processes, the ionic strength
is different on the two sides of the membrane (the inner- and extra-cellular regions).
This difference in ionic strength, in turn, contributes
to an additional gap between the surface
potential on the two sides of the bilayer and plays an important
role, for example, in neuro-transmission processes.

In the present work, we extend the model of Refs.~\cite{BM,WM} in order to mimic
asymmetric biomembranes in a more complete way.
In Sec.~\ref{model}, we introduce a model based on Poisson-Boltzmann theory
to describe phase separation for asymmetric charged bilayers ``sandwiched" between
two ionic solutions having different ionic strength.
In Sec.~\ref{results}, the variation of the critical point and
the entire phase diagram as a function of the ionic strength (Debye length) and
inter-leaflet electrostatic coupling is explored.
It is shown that the phase
separation is enhanced as the salt concentration is increased, and
is due to the enhanced electrostatic screening.
When an electrostatic coupling between the leaflets is introduced,
the lateral phase separation occurs at higher temperatures as compared with the no coupling case.
We also show that the phase-separation temperature increases
when the concentration difference
between the two salt reservoirs becomes larger.
Finally, discussion and comparison with other works are presented in
Sec.~\ref{discussion}.

%%%%%%%%%%%
%  Model  %
%%%%%%%%%%%
\section{Model}
\label{model}

The system is modeled as a binary lipid-bilayer composed
of a mixture of negatively charged  and neutral lipids.
Although  many experiments are done in the presence of added cholesterol
as a third  component,
it is reasonable to stay within the simpler case of a binary lipid mixture.
As cholesterol is non-charged, its presence will not change in any major way our
predictions on the role of electrostatics.

\begin{figure}[t!]
\begin{center}
\includegraphics[scale=0.4]{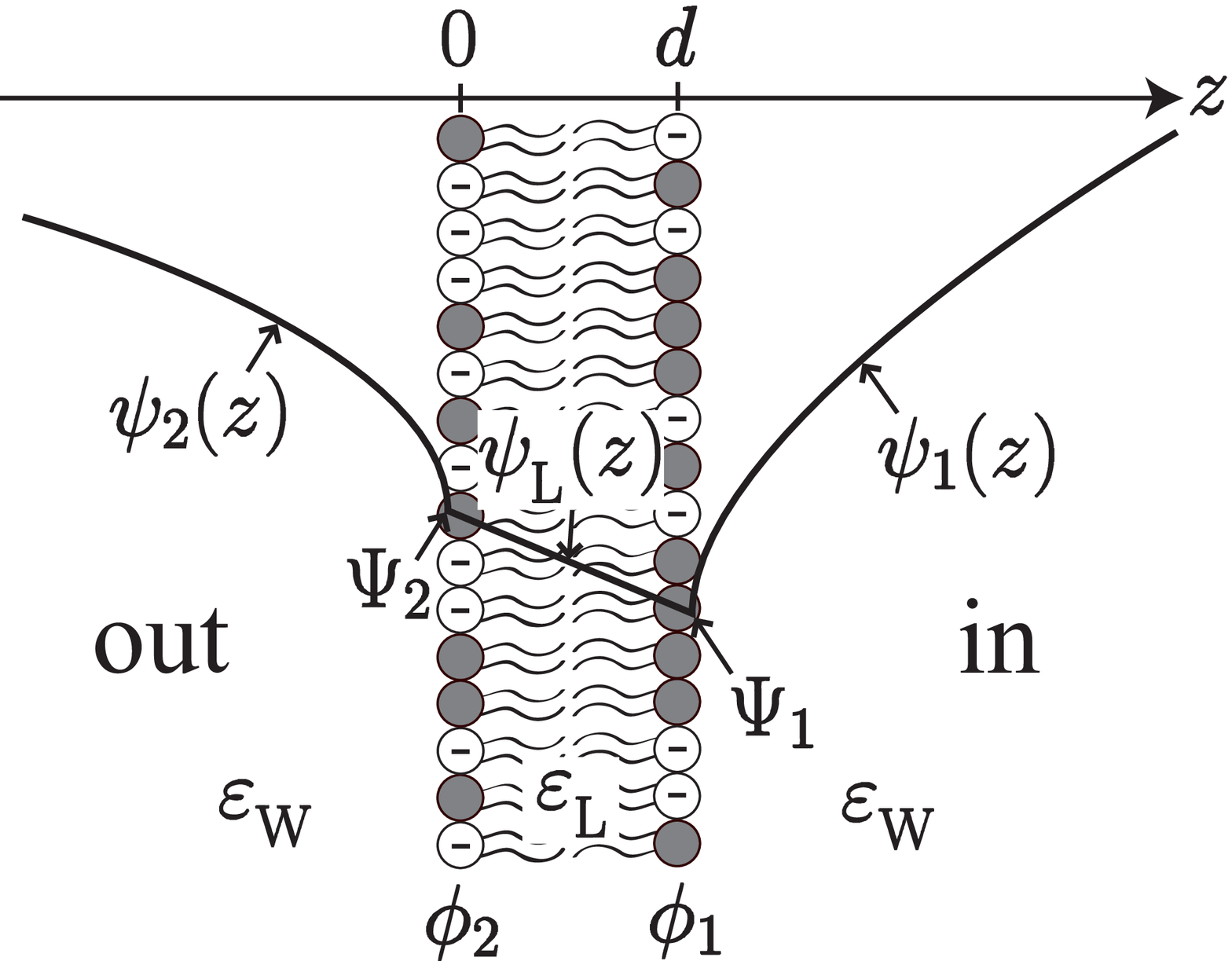}
\end{center}
\caption{\textsf{
Schematic representation of the electrostatic
potential for a binary lipid bilayer consisting of negatively
charged and neutral lipids.
The head groups  of leaflets 1 and 2 are located at the plane $z=d$
and $z=0$, respectively.
The (dimensionless) electrostatic potentials in solutions 1 ($z > d$) and
2 ($z < 0$) are denoted by $\psi_{1}(z)$ and $\psi_{2}(z)$,
respectively.
The dimensionless surface potentials at the two leaflets are denoted by $\Psi_{1}=\psi_{1}(z=d)$
and $\Psi_{2}=\psi_{2}(z=0)$, while
$\psi_{_{\rm L}}(z)$ is the (dimensionless) electric potential inside the hydrocarbon core
region ($0\le z \le d$) of the lipid membrane.
The dielectric constant of the aqueous region (water) and lipid hydrocarbon core are denoted by $\varepsilon_{_{\rm W}}$ and $\varepsilon_{_{\rm L}}$,
respectively.}}
\label{fig1}
\end{figure}

The bilayer consists of two undeformable
and parallel leaflets, lying in the $xy$-plane and in contact, respectively, with the two
monovalent salt solutions, as is shown in Fig.~\ref{fig1}.
Let us label all quantities residing on the inner leaflet by the subscript 1 and those on the outer one by 2.
The inner leaflet (leaflet 1) is located at $z=d$ and is in contact with
the inner reservoir located at $z>d$,
while the outer leaflet (leaflet 2) is located at $z=0$ and is in contact with the outer reservoir
at $z<0$.
In some  experiments spontaneous budding of charged domains toward the vesicular
interior was reported~\cite{SHSY}. This can be explained by considering the increase in the osmotic pressure coupled
with the change in spontaneous curvature of the outer leaflet.
However, in the present study we assume that the
bilayer remains flat and neglect the effect of the osmotic pressure generated across the bilayer.
Our model is further based on three principal assumptions:
(i) Because we do not consider membrane undulations and their curvature, the two leaflets have the same area $A$.
(ii) The neutral and negatively charged lipids have the same cross-sectional area per lipid $\Sigma$.
This is supported by many experiments on fluid-like membranes.
(iii) Furthermore, each leaflet is taken as an incompressible two-dimensional fluid.
Hence, the two leaflets consist of the same total number of lipid molecules $N=A/\Sigma$.

%fig1

The total free energy per lipid molecule is given by
\begin{equation}
\label{free}
f_{\rm tot}(\phi_{1},\phi_{2})=f_{\rm mix}(\phi_{1})+f_{\rm mix}(\phi_{2})+f_{\rm el}(\phi_{1},\phi_{2}),
\end{equation}
where all energies are measured in units of $k_{\rm B}T$ ($k_{\rm B}$ is the Boltzmann constant and $T$ the temperature),
and $\phi_{1}$ and $\phi_{2}$ are the mole fractions of the
negatively charged lipid in leaflets 1 and 2, respectively.
The free energy $f_{\rm tot}$ consists of two non-electrostatic $f_{\rm mix}$ terms (one for each leaflet),
and an electrostatic one $f_{\rm el}$ that induces an inter-leaflet coupling.

The non-electrostatic $f_{\rm mix}$ is the
Flory-Huggins free energy of lateral mixing in
each of the leaflets separately
\begin{equation}
\label{FH}
f_{\rm mix}(\phi)= \phi \ln \phi + (1-\phi) \ln (1-\phi) + \chi \phi (1-\phi),
\end{equation}
where $\chi$ is the (non-electrostatic) interaction parameter between the two different lipids.
Quite generally, $\chi$ is  taken to vary inversely with the temperature.
The free energy $f_{\rm mix}$ is a sum of the ideal entropy and enthalpy of mixing between the
two lipid species.
In the absence of the electrostatic interaction, $f_{\rm el}=0$, the two-phase coexistence region in
the ($\chi$,$\phi$)-plane is delimited by a demixing curve that terminates at the
critical point: $\chi_{\rm c}=2$ and $\phi_{\rm c}=0.5$.

The electrostatic free energy $f_{\rm el}$ can be calculated through
the charging process~\cite{VO}:
\begin{equation}
\label{charging}
f_{\rm el}(\phi_{1},\phi_{2})
=-\int_{0}^{\phi_{1}}\Psi_{1}(s_{1},0) \,
{\rm d}s_{1}
-\int_{0}^{\phi_{2}}\Psi_{2}(\phi_{1},s_{2}) \,
{\rm d} s_{2},
\end{equation}
where $\Psi_{1}=e\Phi_1(z{=}d)/k_{\rm B}T$ and $\Psi_{2}=e\Phi_2(z{=}0)/k_{\rm B}T$ are the dimensionless
surface potentials on
leaflets 1 ($z=d$) and 2 ($z=0$), respectively, and $e$ is the elementary charge, while
the electric potential inside the lipid hydrocarbon
core region ($0\le z \le d$) is denoted by $\Phi_{_{\rm L}}(z)$.
Within mean-field theory, the electric potential $\Phi(z)$ satisfies the Poisson-Boltzmann equation
\begin{equation}
\label{PB}
\frac{{\rm d}^{2} \Phi}{{\rm d} z^{2}}=\frac{2en}{\varepsilon_{_{\rm W}}} \sinh \frac{e \Phi}{k_{\rm B}T},
\end{equation}
where $n$ is the salt concentration in bulk and $\varepsilon_{_{\rm W}}$
the dielectric constant of the aqueous solution (water).
Using the dimensionless electrostatic potential $\psi(z) \equiv e \Phi(z)/k_{\rm B}T$, we obtain the following
Poisson-Boltzmann equations in regions 1 and 2 for $\psi_{1,2}(z)$, and Laplace equation in the core-region
for $\psi_{_{\rm L}}(z)$;
\begin{eqnarray}
\label{dim_PB}
\frac{{\rm d}^{2} \psi_1}{{\rm d} z^{2}} &=& \kappa_{\rm 1}^{2} \sinh \psi_1 ~~~~~ z>d, \nonumber \\
\frac{{\rm d}^{2} \psi_2}{{\rm d} z^{2}} &=& \kappa_{\rm 2}^{2} \sinh \psi_2 ~~~~~ z<0, \nonumber \\
\frac{{\rm d}^{2} \psi_{_{\rm L}}}{{\rm d} z^{2}}&=&0  ~~~~~~~~~~~~~~~0\le z \le d,
\end{eqnarray}
where $\kappa_{\rm 1}^{-1}$ and $\kappa_{\rm 2}^{-1}$ are the Debye screening length
in regions 1 and 2, respectively, defined by $(\kappa_{1,2})^{-1}=\sqrt{\varepsilon_{_{\rm W}}k_{\rm B}T/2e^{2}n_{1,2}}$.
Using the Gauss theorem, the boundary conditions of the two leaflets
are given by
\begin{eqnarray}
\label{Gauss}
\left. \varepsilon_{_{\rm W}} \frac{{\rm d} \psi_1}{{\rm d} z} \right|_{z=d} - \left.
\varepsilon_{_{\rm L}} \frac{{\rm d} \psi_{_{\rm L}}}{{\rm d} z} \right|_{z=d} &=& \frac{e^2\phi_{1}}{ k_{\rm B}T\Sigma}, \nonumber \\
\left. \varepsilon_{_{\rm W}} \frac{{\rm d} \psi_2}{{\rm d} z}
\right|_{z=0} - \left. \varepsilon_{_{\rm L}} \frac{{\rm d} \psi_{_{\rm L}}}{{\rm d} z} \right|_{z=0} &=& -\frac{e^2\phi_{2}}{ k_{\rm B}T\Sigma},
\end{eqnarray}
where $\Sigma$ is the cross-sectional area per lipid head group and $\varepsilon_{_{\rm L}}$ is the dielectric constant of
the lipid hydrocarbon core-region.
Far from both sides of the membrane, the bulk electric potential is taken to be zero,
$\psi_{1}(z \rightarrow \infty)= 0$
and $\psi_{2}(z \rightarrow -\infty)=0$.

Under these boundary conditions, the above equations
can be solved analytically, yielding transcendental expressions for the two surface potentials, $\Psi_1$ and $\Psi_2$
\begin{eqnarray}
\label{surface12}
\Psi_{1} &=& -2 \sinh^{-1} \left(p_{1} \phi_{1} + p_{1} H \Delta \Psi \right) \, \nonumber \\
\Psi_{2} &=& -2 \sinh^{-1} \left(p_{2} \phi_{2} - p_{2} H \Delta \Psi \right)\,
\end{eqnarray}
where $\Delta \Psi = \Psi_1 - \Psi_2$ is the membrane potential gap. Throughout the paper we will make use of three dimensionless quantities
defined as:
\begin{eqnarray}
\label{p_and_H}
p_1 &=& \frac{2\pi \ell_{\rm B}}{\kappa_1\Sigma} ~~\sim ~~\kappa_1^{-1}, \nonumber \\
p_2 &=& \frac{2\pi\ell_{\rm B}}{\kappa_2\Sigma} ~~\sim ~~\kappa_2^{-1}, \nonumber \\
H   &=& \frac{\Sigma}{4 \pi \varepsilon_{_{\rm W}}\ell_{\rm B}} \frac{{\varepsilon}_{_{\rm L}}}{d}~~\sim ~~\frac{\varepsilon_{_{\rm L}}}{d}\, ,
\end{eqnarray}
where $\ell_{\rm B} = e^2/(4\pi \varepsilon_{_{\rm W}} k_{\rm B} T)\simeq 7$\AA\ is the Bjerrum length.
The ratio $p_{2}/p_{1}=\kappa_1/\kappa_2$ is related to the
ratio between the two Debye screening lengths, $(\kappa_{1,2})^{-1}$, of the two solutions,
and $H$ is the rescaled electrostatic coupling parameter between the two leaflets.
By further defining $\overline{\phi}_{1}\equiv \phi_{1}+ H \Delta \Psi$ and $\overline{\phi}_{2}\equiv \phi_{2}- H \Delta \Psi$
as effective mole fractions of the negatively charged lipid in each leaflet, we can rewrite Eq.~(\ref{surface12}) as:
\begin{eqnarray}
\label{surface3}
\Psi_{1}&=& -2 \sinh^{-1} (p_{1} \overline{\phi}_{1})\, , \nonumber \\
\Psi_{2}&=& -2 \sinh^{-1} (p_{2} \overline{\phi}_{2})\, .
\end{eqnarray}
Furthermore, the potential difference across the membrane $\Delta \Psi=\Psi_{1}-\Psi_{2}$
satisfies a transcendental equation
\begin{equation}
\label{difference}
\Delta \Psi = -2 \sinh^{-1} (p_{1}
\overline{\phi}_{1}) + 2 \sinh^{-1} (p_{2} \overline{\phi}_{2})\, ,
\end{equation}
since $\overline{\phi}_{1,2}$ themselves depend on $\Delta \Psi$ as defined above.

By substituting back the surface potentials, Eq.~(\ref{surface12}),
into Eq.~(\ref{charging}), we obtain the final expression of the electrostatic free energy
\begin{equation}
\label{electro_free}
f_{\rm el}(\phi_{1},\phi_{2})=\frac{1}{2}H (\Delta \Psi)^{2} +
f_{\rm m}(\overline{\phi}_{1}) + f_{\rm m}(\overline{\phi}_{2}).
\end{equation}
The first term corresponds to the charging energy stored in the
two charged leaflets that are analogous to a two-plate capacitor.
The second and third terms represent the electrostatic energy of the two
isolated charged monolayers.
In the absence of the electrostatic coupling between the two leaflets
($H=0$), $f_{\rm el}$ is simply the sum of the
free energies of two isolated leaflets, i.e.,
$f_{\rm el}(\phi_{1},\phi_{2})=f_{\rm m}(\phi_{1})+f_{\rm m}(\phi_{2})$.
Within the Poisson-Boltzmann theory, $f_{\rm m}$ was derived
in Ref.~\cite{Colloid}, and we briefly repeat this derivation in the Appendix.
The result is
\begin{equation}
\begin{split}
\label{isolated}
f_{\rm m}(\overline{\phi}) &= 2 \int_{0}^{\overline{\phi}} \sinh^{-1}(p s)
\, {\rm d} s \\
&= 2 \overline{\phi} \left[ \frac{1-\sqrt{(p \overline{\phi})^{2}+1}}
{p  \overline{\phi}} + \ln \left( p \overline{\phi} +
\sqrt{(p \overline{\phi})^{2}+1} \right) \right].
\end{split}
\end{equation}
$p=p_{1}$ for $\overline{\phi}=\overline{\phi}_{1}$, and $p=p_{2}$ for $\overline{\phi}=\overline{\phi}_{2}$.
We first solve Eq.~(\ref{difference}) to obtain $\Delta \Psi$, and then use it to
calculate the total free energy from Eqs.~(\ref{free}), (\ref{FH}),
(\ref{electro_free}) and (\ref{isolated}).

In order to analyze the membrane stability toward a lateral phase separation, we
calculate the spinodal surface given in the ($\phi_1$, $\phi_2$, $\chi$) parameter
space by the condition
\begin{equation}
\label{spinodal1}
\left( \frac{\partial^{2} f_{\rm tot}}{\partial \phi_{1}^{2}} \right)
\left( \frac{\partial^{2} f_{\rm tot}}{\partial \phi_{2}^{2}} \right)
- \left( \frac{\partial^{2} f_{\rm tot}}{\partial \phi_{1} \partial \phi_{2}} \right)^{2} = 0.
\end{equation}
Using the total free-energy of Eq.~(\ref{free}), we obtain
\begin{equation}
\begin{split}
\label{spinodal2}
\left[ \frac{1}{\phi_{1} (1-\phi_{1})} - 2 \chi + \left( \frac{\partial^{2} f_{\rm el}}{\partial \phi_{1}^{2}} \right) \right] & \\
\left[ \frac{1}{\phi_{2} (1-\phi_{2})} - 2 \chi + \left( \frac{\partial^{2} f_{\rm el}}{\partial \phi_{2}^{2}} \right) \right] 
& = \left( \frac{\partial^{2} f_{\rm el}}{\partial \phi_{1} \partial \phi_{2}} \right)^{2}.
\end{split}
\end{equation}
The first and second derivatives of the electrostatic free-energy, $f_{\rm el}(\phi_{1},\phi_{2})$, are given by
Eq.~(\ref{electro_free})
and can be expressed as
\begin{eqnarray}
\label{first}
 \frac{\partial f_{\rm el}}{\partial \phi_{1}} =2 \,\sinh^{-1}(p_{1} \overline{\phi}_{1}) &=& -\Psi_{1}, \nonumber \\
\frac{\partial f_{\rm el}}{\partial \phi_{2}} =2 \,\sinh^{-1}(p_{2} \overline{\phi}_{2}) &=& -\Psi_{2},
\end{eqnarray}
and
\begin{eqnarray}
\label{second}
 \frac{\partial^{2} f_{\rm el}}{\partial \phi_{1}^{2}} &=& \frac{f_{\rm m}''(\overline{\phi}_{1})+ H f_{\rm m}''(\overline{\phi}_{1}) f_{\rm m}''(\overline{\phi}_{2})}
{1+H [f_{\rm m}''(\overline{\phi}_{1})+f_{\rm m}''(\overline{\phi}_{2})]}, \nonumber \\
\frac{\partial^{2} f_{\rm el}}{\partial \phi_{2}^{2}} &=& \frac{f_{\rm m}''(\overline{\phi}_{2})+ H f_{\rm m}''(\overline{\phi}_{1}) f_{\rm m}''(\overline{\phi}_{2})}
{1+H [f_{\rm m}''(\overline{\phi}_{1})+f_{\rm m}''(\overline{\phi}_{2})]}, \nonumber \\
\frac{\partial^{2} f_{\rm el}}{\partial \phi_{1} \partial \phi_{2}} &=& \frac{H f_{\rm m}''(\overline{\phi}_{1}) f_{\rm m}''(\overline{\phi}_{2})}
{1+H [f_{\rm m}''(\overline{\phi}_{1})+f_{\rm m}''(\overline{\phi}_{2})]},
\end{eqnarray}
where $f_{\rm m}''(\overline{\phi})={\rm d}^{2} f_{\rm m}(\overline{\phi})/
{\rm d} \overline{\phi}^{2} =2p/\sqrt{(p\overline{\phi})^{2}+1}$.
Finally, the spinodal surface $\chi_{\rm sp}(\phi_{1},\phi_{2})$ for fixed $p_1, p_2, H$
can be determined according to Eqs.~(\ref{difference}), (\ref{spinodal2})
and (\ref{second}).

The chemical potential is a function of the lipid concentration
\begin{equation}
\begin{split}
\label{chemical}
\mu_{1}&=\frac{\partial f_{\rm tot}}{\partial \phi_{1}}=2 \sinh^{-1} (p \overline{\phi}_{1}) + \ln \frac{\phi_{1}}{1-\phi_{1}} +\chi (1-2 \phi_{1}), \\
\mu_{2}&=\frac{\partial f_{\rm tot}}{\partial \phi_{2}}=2 \sinh^{-1} (p \overline{\phi}_{2}) + \ln \frac{\phi_{2}}{1-\phi_{2}} +\chi (1-2 \phi_{2}).
\end{split}
\end{equation}
Under the assumptions mentioned in the beginning in this section, the condition for phase coexistence (the binodal line) requires the equality of the
chemical potential of the two phases, A and B, on each of the two leaflets,
meaning that both $\mu_{\rm 1,A}=\mu_{\rm 1,B}$ and
$\mu_{\rm 2,A}=\mu_{\rm 2,B}$ should be  simultaneously satisfied on the two leaflets.
In addition, the thermodynamic potential
\begin{equation}
\label{thermodynamic}
g(\phi_{1},\phi_{2}) = f_{\rm tot}(\phi_{1},\phi_{2}) - \mu_{1} \phi_{1} - \mu_{2} \phi_{2}
\end{equation}
should be equal for the two phases, i.e, $g_{\rm A}=g_{\rm B}$.
These conditions are sufficient to calculate the two-phase coexistence region
as well as the corresponding tie-lines inside the coexistence region, as will be discussed below in
Sec.~\ref{results}. We remark that the same conditions were also used in previous works~\cite{WLM, PS2}.

%%%%%%%%%%%%
\subsection{The symmetric in/out reservoir case, $p_1=p_2$}
%%%%%%%%%%%%%

Let us first discuss the case of equal ionic strength ($p_{1}=p_{2}$) of the two solutions
and review some of the results reported already in Ref.~\cite{BM}.
The critical point, because of the in and out symmetry, is located on
the $\phi_{1}=\phi_{2}$ line, and
the expression for the spinodal line is given by
\begin{equation}
\label{symmetry}
\chi_{\rm sp}=\frac{1}{2 \phi (1- \phi)}+\frac{p}{\sqrt{1 + p^{2} \phi^{2}}+4 p H},
\end{equation}
where $p=p_{1}=p_{2}$ and $\phi=\phi_{1}=\phi_{2}$.
The minimum value of $\chi_{\rm sp}$ corresponds to the critical point.

Let us take now explicitly the limits of high and low salt concentrations
for the symmetric $p$ case.
The high-salt  limit ($p \ll 1$) corresponds to the
Debye-H\"uckel regime for which the spinodal line is given by
\begin{equation}
\label{high_salt_sp}
\chi_{\rm sp} \approx \frac{1}{2\phi(1-\phi)}+\frac{p}{1+4pH}\, ,
\end{equation}
and the critical point is located at
$\chi_{\rm c}=2, \phi_{\rm c}=0.5$.
This means that the bilayer behaves as if it were a neutral membrane
when the salt concentration is sufficiently high.

In the other limit of low salt ($p \gg 1$), the spinodal line is
given by
\begin{equation}
\label{low_salt_sp}
\chi_{\rm sp} \approx \frac{1}{2\phi(1-\phi)}+\frac{1}{\phi+4H},
\end{equation}
and it is independent of $p$.
If the electrostatic coupling is weak enough ($H \ll 1$),
the respective critical point is located at
\begin{eqnarray}
\label{low_salt}
\chi_{\rm c} &=& (2+\sqrt{3})(1-\frac{8H}{3}), \nonumber \\
\phi_{\rm c} &=& \frac{3-\sqrt{3}}{2}-\frac{4H(\sqrt{3}-1)}{3}.
\end{eqnarray}
In the absence of the electrostatic coupling ($H=0$), the critical point
coincides with that in the previous works~\cite{BM,WM,GB}:
\begin{eqnarray}
\chi_{\rm c}&=&2+\sqrt{3} \simeq 3.73\, ,\nonumber\\
\phi_{\rm c}&=&(3-\sqrt{3})/2 \simeq 0.634\, .
\end{eqnarray}

%%%%%%%%%%%%%
%  Results  %
%%%%%%%%%%%%%
\section{Global phase diagrams and critical behavior}
\label{results}

%fig2
\begin{figure}[t!]
\begin{center}
\includegraphics[scale=0.65]{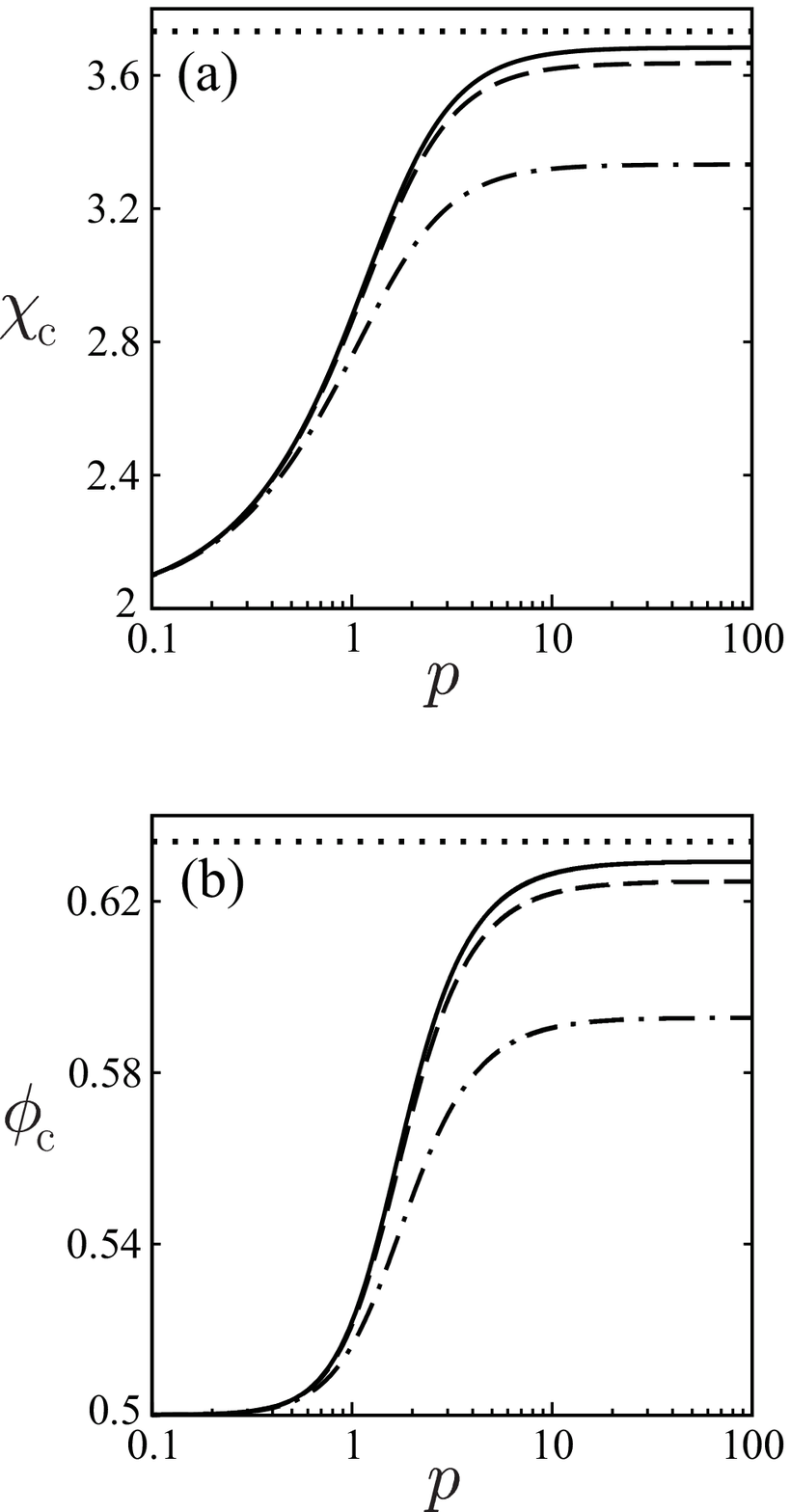}
\end{center}
\caption{\textsf{
(a) The critical interaction $\chi_{\rm c}$,  and (b) critical composition $\phi_{\rm c}$
as a function of $p$ on a semi-log plot for different $H$ values in the
symmetric case ($p_{1}=p_{2}=p$).
The solid, dashed, and dot-dashed lines are for $H=0.005, 0.01, 0.05$,
respectively.
For $H=0$, the critical values are represented by the dotted lines located at
(a) $\chi_{\rm c}=2+\sqrt{3} \simeq3.73$ and (b) $\phi_{\rm c}=(3-\sqrt{3})/2 \simeq 0.634$.
}}
\label{fig2}
\end{figure}

Based on the model described in the previous section, we present results
for the critical point and  phase diagrams in the case of two coupled and charged leaflets.
In particular, we focus on the difference between a symmetric case
in which the two solutions have the same ionic strength ($p_{1}=p_{2}$)
and the asymmetric case in which they differ
($p_{1} \neq p_{2}$).

We start by giving some
estimates for the physical choices of the parameter values of $p_{1}$, $p_{2}$, and $H$.
The cross-sectional area of a lipid headgroup is set to about $\Sigma\simeq 65$\,\AA $^{2}$
(taken to be the same for the two lipids), and
the membrane thickness is $d\simeq 50$\,\AA.
From these values together with $\varepsilon_{_{\rm W}} \simeq 80 \varepsilon_{0}$,
$\varepsilon_{_{\rm L}} \simeq 3 \varepsilon_{0}$ ($\varepsilon_{0}$ is the vacuum
permittivity) and $T\simeq 300$\,K, we obtain
$p \simeq 0.67  \kappa^{-1}$ where
$\kappa^{-1}$ is measured in Angstroms, and $H=5.6 \times 10^{-4}$.
In the following, we use a range of values
$p=0.5, 5, 50$ and extended the $H$ range to cover $H=0.005, 0.01, 0.05$.
In the case of monovalent salt,
the values $p=0.5, 5, 50$ correspond, respectively, to
$\kappa^{-1}=0.75, 7.5, 75$\,\AA\  or to
$n=16$\,M, $160$\,mM, $1.6$\,mM.
Since the electrostatic coupling parameter $H$ is proportional to $\varepsilon_{_{\rm L}}/d$,
the larger $H$ values correspond to smaller $d$ and/or larger $\varepsilon_{_{\rm L}}$.
Obviously, the range in the $p$ parameter of $10^{2}$
gives a range of $10^{4}$ in ionic strength, which probably is too large to compare with experiments. Hence
it should be taken just to
indicate trends with changing the salinity.

We show first the phase separation in asymmetric charged bilayers
sandwiched between the two ionic solutions having the {\it same}
ionic strength ($p_1=p_2$), repeating the results of Ref.~\cite{BM}.
Then, we present results for asymmetric charged leaflets that are
in contact with two ionic solutions having {\it different} ionic strengths ($p_1\ne p_2$).

%%%%%%%%%%%%%%%%%%%%%%%%%%%%%%%%%%%%%%%%%%%%%%%%%%
\subsection{Symmetric in/out reservoirs: $p_1=p_2$}
%%%%%%%%%%%%%%%%%%%%%%%%%%%%%%%%%%%%%%%%%%%%%%%%%%%

In Fig.~\ref{fig2}, we show the critical interaction
and  composition ($\chi_{\rm c}, \phi_{\rm c}$) as a function of $p$
for the symmetric reservoir case, $p=p_{1}=p_{2}$.
Both $\chi_{\rm c}$ and $\phi_{\rm c}$ are obtained by numerically
minimizing Eq.~(\ref{symmetry}). The solid, dashed, dot-dashed lines correspond, respectively, to $H=0.005,
0.01,$ and 0.05.
The horizontal dotted lines located at $\chi_{\rm c}=2+\sqrt{3}=3.73$
and $\phi_{\rm c}=(3-\sqrt{3})/2=0.634$  are the critical values
for the $H \rightarrow 0$ limit and for high-salt conditions.

When $\chi > \chi_{\rm c}$, the membrane undergoes a lateral phase separation,
whereas for $\chi < \chi_{\rm c}$ the membrane is in a single (homogeneous) phase.
For all $H$ values, we find that $\chi_{\rm c}$ becomes smaller as $p$
decreases.
Namely, the phase-separation temperature is increased
and the two-phase region has a larger extent for smaller $p$.
This is because the Coulombic repulsion between the charged lipids
becomes too weak to overpower the attractive interaction that drives the
phase separation.
When $H$ is large, $\chi_{\rm c}$ becomes much smaller, as
will be discussed later in Sec.~\ref{discussion}.
In the high salt limit or when $p$ is small enough,
the limit of $\chi_{\rm c}\to 2$ and $\phi_{\rm c}\to 0.5$ can be seen,
as was discussed above in Sec.~\ref{model}.
In the opposite low salt limit, $\chi_{\rm c}$ and $\phi_{\rm c}$
approaches those of Eq.~(\ref{low_salt}).

%%%%%%%%%%%%%%%%%%%%%%%%%%%%%%%%%%%%%%%%%%%%%%%%%%%%%%%%%
\subsection{Asymmetric in/out reservoirs: $p_1\ne p_2$}
%%%%%%%%%%%%%%%%%%%%%%%%%%%%%%%%%%%%%%%%%%%%%%%%%%%%%%%%%

%fig3
\begin{figure*}[t!]
\begin{center}
\includegraphics[scale=0.75]{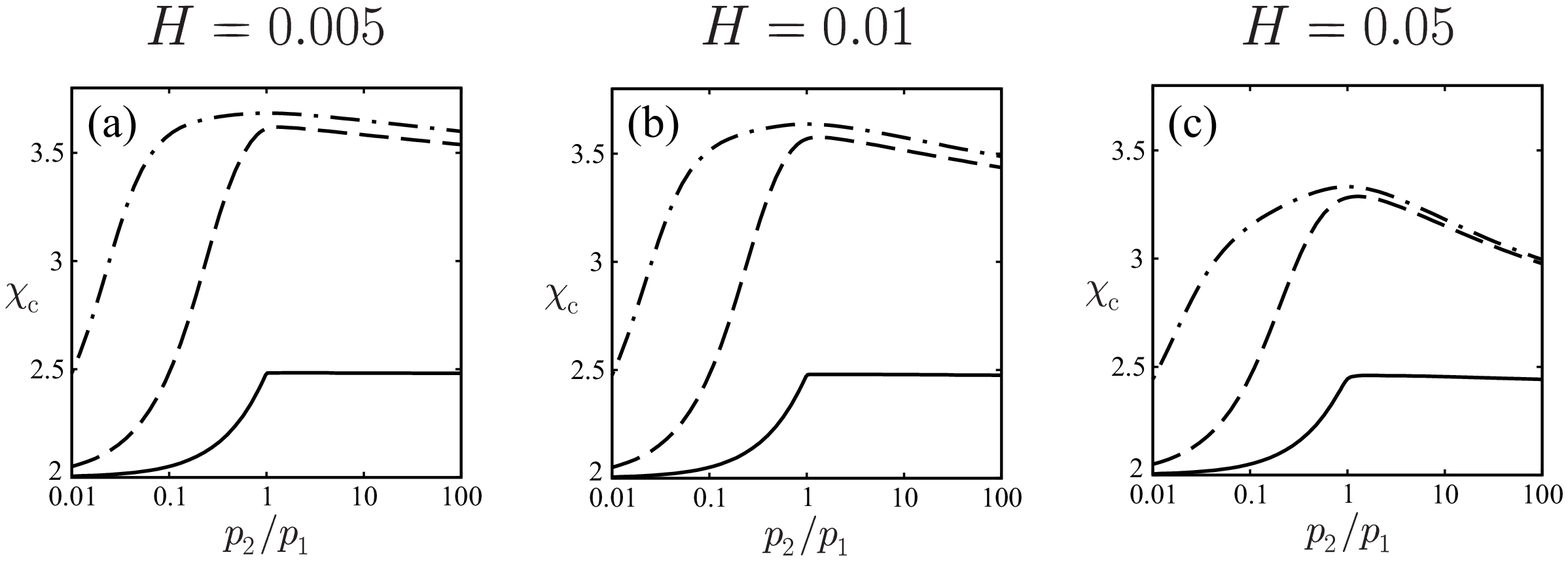}
\end{center}
\caption{\textsf{
The critical value $\chi_{\rm c}$ as function of the asymmetry parameter, $p_{2}/p_{1}$,
which also is equal to  the ratio of the two Debye lengths, $\kappa_1/\kappa_2$.
The solid, dashed, dot-dashed lines are for $p_{1}=0.5, 5, 50$,
respectively.
The values of the $H$ parameter are: (a) $H=0.005$, (b) $H=0.01$, and
(c) $H=0.05$.
}}
\label{fig3}
\end{figure*}

The calculated value of $\chi_{\rm c}$ for the asymmetric case, in which
the two solutions have different ionic strengths, is plotted in
Fig.~\ref{fig3}.
Since $\chi_{\rm sp}$ equals to $\chi_{\rm c}$ at the critical point,
$\chi_{\rm c}$ is obtained by calculating numerically
the minimum value of $\chi_{\rm sp}$ via Eq.~(\ref{spinodal2}).
The horizontal axis is the ratio $p_{2}/p_{1}$ in logarithmic scale,
where $p_{2}/p_{1}=1$ corresponds to equal ionic strength of the two reservoirs.
The solid, dashed, and dot-dashed lines denote fixed $p_{1}=0.5$, 5, and 50,
respectively, while the coupling parameter $H$ has three different values
$H=0.005, 0.01$, and 0.05 in the three figure parts.
Similar to the findings of Fig.~\ref{fig2}, $\chi_{\rm c}$ decreases as the coupling $H$ increases.
This dependence of $\chi_{\rm c}$ on $H$ is similar as in the symmetric case
and is further discussed in Sec.~\ref{discussion}.
In all plotted cases there is a peak around
$p_{2}/p_{1}=1$, without a  break in its slope there even for the solid line.
For $p_{2}/p_{1}<1$, $\chi_{\rm c}$ decreases as the salt concentration
in solution 2 is increased, similar to the symmetric case.

Moreover, the Coulombic repulsion between charged lipids in the
same leaflet becomes weak and $\chi_{\rm c}$
decreases when $p_{2}/p_{1}$ becomes smaller.
For $p_{2}/p_{1}>1$, $\chi_{\rm c}$ is found to decrease
and its value is more pronounced when $H$ becomes larger, although the electrostatic screening is weaker.
The physical origin of this behavior will be further discussed in
Sec.~\ref{discussion}.

\begin{figure*}[t!]
\begin{center}
\includegraphics[scale=0.65]{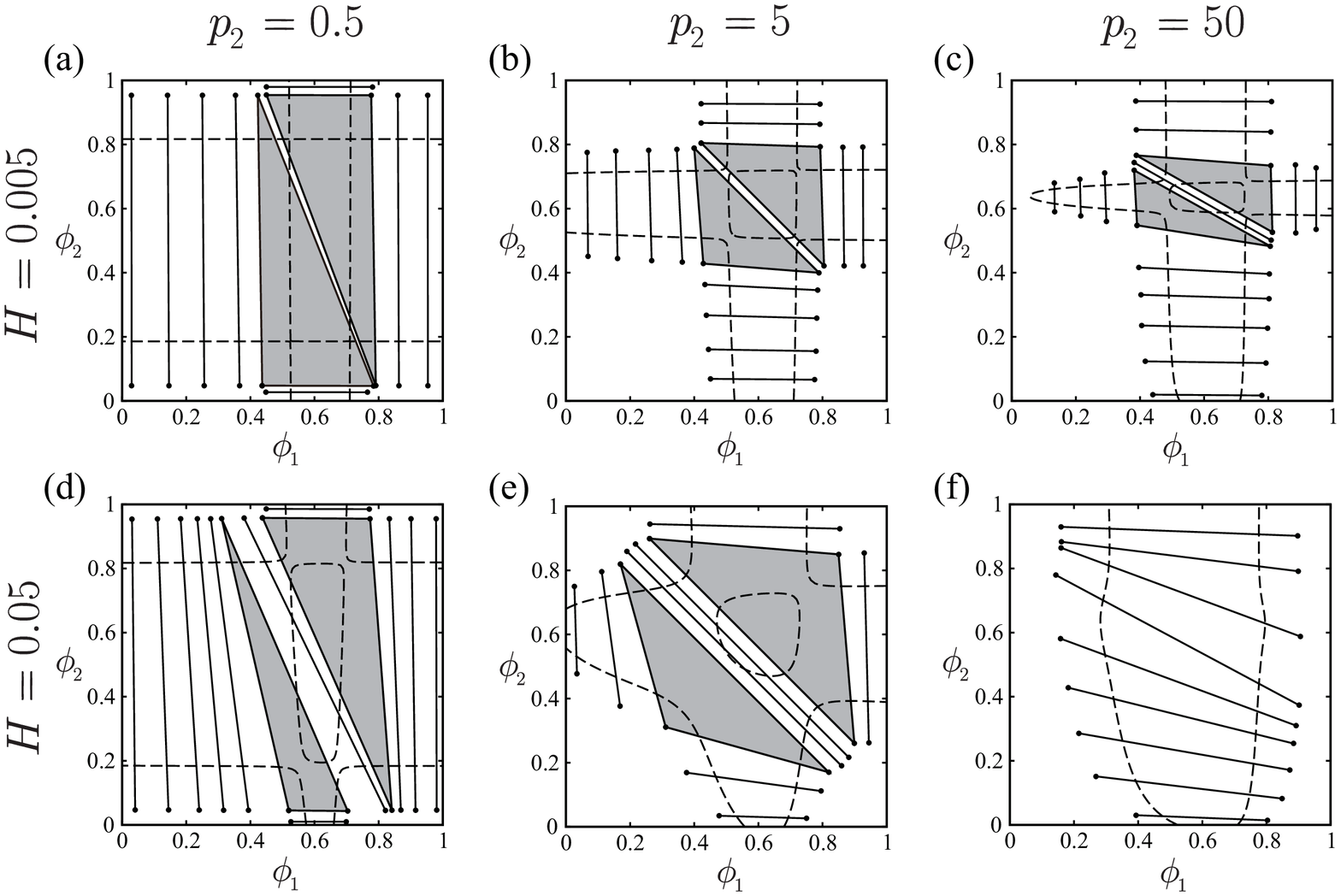}
\end{center}
\caption{\textsf{
Phase diagrams in the ($\phi_{1}, \phi_{2}$) plane for $\chi=3.8$
and $p_{1}=5$.
The solid lines are the tie-lines and the dashed lines are the
spinodal lines.
The  three-phase-coexistence regions are denoted by shaded triangles.
The values of the other parameters are:
 (a) $H=0.005$ and $p_{2}=0.5$; (b) $H=0.005$ and $p_{2}=5$;
(c) $H=0.005$ and $p_{2}=50$; (d) $H=0.05$ and $p_{2}=0.5$;
(e) $H=0.05$ and $p_{2}=5$; and (f) $H=0.05$ and $p_{2}=50$.
}}
\label{fig4}
\end{figure*}

\begin{figure*}[t!]
\begin{center}
\includegraphics[scale=0.65]{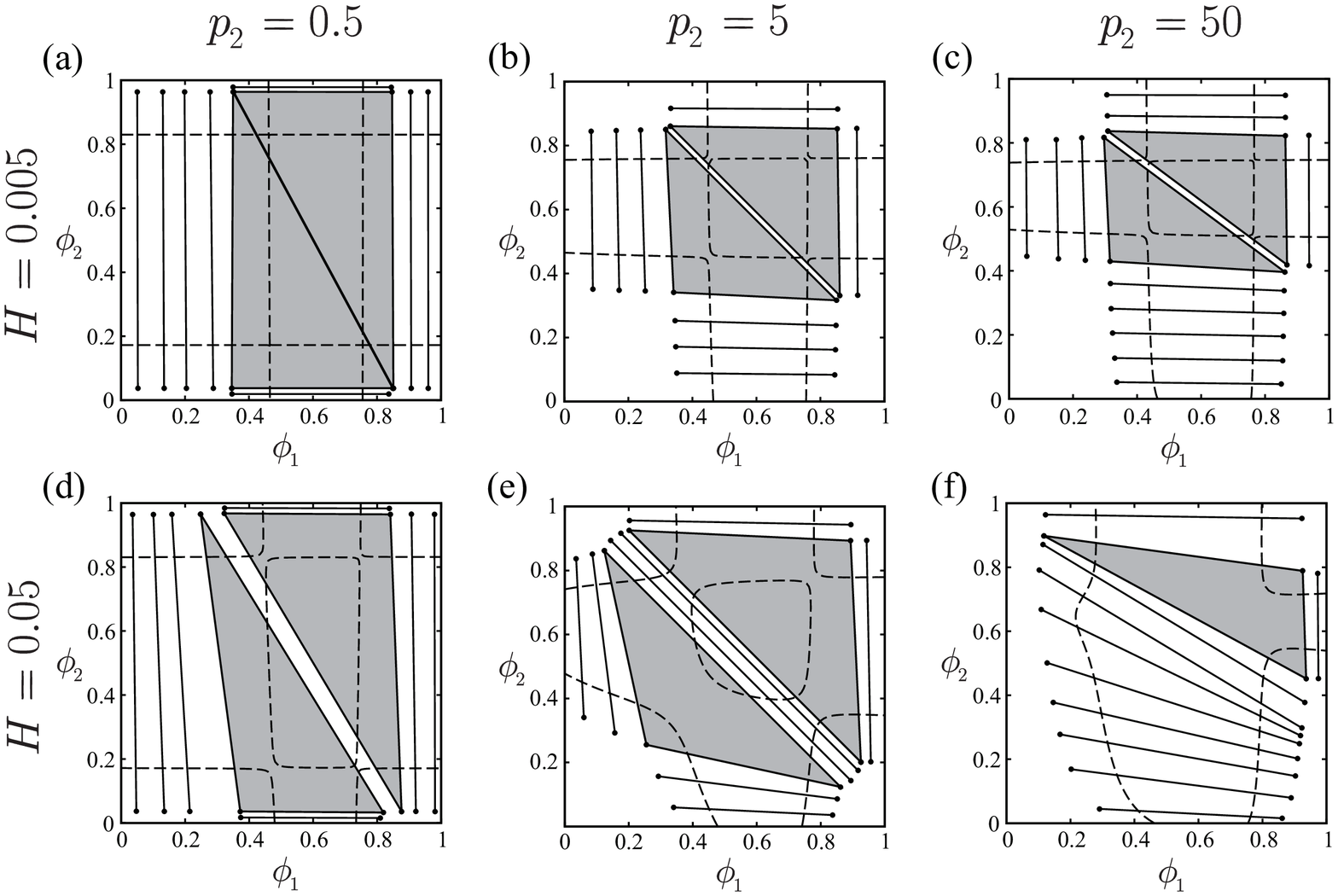}
\end{center}
\caption{\textsf{
Phase diagrams in the ($\phi_{1}, \phi_{2}$) plane for $\chi=4.0$
and $p_{1}=5$.
The solid lines are the tie-lines and the dashed lines are the
spinodal lines.
The  three-phase-coexistence regions are denoted by shaded triangles.
The values of the other parameters are:
 (a) $H=0.005$ and $p_{2}=0.5$; (b) $H=0.005$ and $p_{2}=5$;
(c) $H=0.005$ and $p_{2}=50$; (d) $H=0.05$ and $p_{2}=0.5$;
(e) $H=0.05$ and $p_{2}=5$; and (f) $H=0.05$ and $p_{2}=50$.
}}
\label{fig5}
\end{figure*}

The calculated phase diagrams in the ($\phi_{1},\phi_{2}$) plane
for $\chi=3.8$ and $4.0$ are shown in Figs.~\ref{fig4} and \ref{fig5},
respectively.
We plot  all phase diagrams for a fixed $p_{1}=5$, while varying
$p_{2}$ and $H$.
As can be seen in Fig.~\ref{fig3}, the system undergoes a phase separation for $\chi$ values larger than $\chi_{\rm c}$.
The solid lines are the tie-lines within the coexistence region,
while the dashed lines correspond to
the spinodal lines.
Each pair of points connected by the tie-line satisfies the conditions
$\mu_{\rm 1,A}=\mu_{\rm 1,B}$, $\mu_{\rm 2,A}=\mu_{\rm 2,B}$ [Eq.~(\ref{chemical})] and
$g_{\rm A}=g_{\rm B}$ [Eq.~(\ref{thermodynamic})] while
the spinodal lines are obtained from Eqs.~(\ref{difference}), (\ref{spinodal2}) and
(\ref{second}).
In some of the figures, inner and external spinodal lines can be seen. However, the inner lines
are preempted by the external ones.
Since Fig.~\ref{fig4}(b), (e) and Fig.~\ref{fig5}(b), (e) are for the
symmetric case ($p_1=p_2$), these four phase diagrams are symmetric
with respect to the diagonal line, $\phi_{1}=\phi_{2}$.
All other phase diagrams correspond to asymmetric situations,
$p_{1} \neq p_{2}$.

The phase-coexistence region becomes relatively larger when $p_{2}/ p_{1}\ll 1$,
corresponding to strongly screened systems.
However, $\chi_{\rm c}$ for the asymmetric case is smaller than that
of the symmetric case.
When $H=0.005$, most of the tie-lines are nearly parallel to either
$\phi_{1}$- or $\phi_{2}$-axis.
This implies that the phase separation in each leaflet takes place
almost independently and without any noticeable
correlation with the second leaflet.
On the other hand, for $H=0.05$, the tie-lines are tilted, indicating
a strong coupling between the two leaflets.

The three-phase-coexistence regions are indicated by shaded triangles in
the phase diagrams.
The tie-lines in the vicinity of the three-phase-coexistence regions
are almost orthogonal to the principal diagonal $\phi_{1}=\phi_{2}$, indicating
that $\phi_{1}$ is small when $\phi_{2}$ is large and vice versa. This
occurs when the local concentration of charged lipid in one
of the leaflets is increased, while the concentration in the second
leaflet is decreased conversely.
This behavior is caused by the electrostatic coupling between the leaflets,
as is discussed in the next section.

%fig4

%fig5

%%%%%%%%%%%%%%%%%%%%%%%%%%%%
%  Summary and Discussion  %
%%%%%%%%%%%%%%%%%%%%%%%%%%%%
\section{Discussion and Comparison with Experiments}
\label{discussion}

Several points merit further discussion.
First, we compare our theoretical results with experiments, where it was
reported that the phase separation is enhanced
when salt is added to charged bilayers~\cite{SHSY}.
Such a tendency is well reproduced in Figs.~\ref{fig4} and \ref{fig5}
when the salt concentration in solution 2 is increased or equivalently $p_{2}$ is
decreased. A similar trend was also presented in previous works~\cite{MHB,HMB,MBM} for charged
membranes without any electrostatic inter-leaflet coupling.
In addition, it was reported by Vequi-Suplicy \textit{et al.}~\cite{Dimova} that
the phase-separation temperature increases in the presence of
salt.

In previous experimental studies~\cite{SHSY,Dimova},  vesicles were first
dispersed in the aqueous solution,
and only then salt was added, affecting
only the ionic strength of the solution outside the vesicles.
For this experimental procedure the value of
$p_{2}/p_{1}$ is smaller than unity (recalling that $p_2$ corresponds to the outer reservoir).
This situation is presented in Fig.~\ref{fig3} where
$\chi_{\rm c}\sim 1/T_{\rm c}$ decreases as $p_{2}/p_{1}<1$ becomes smaller,
in accord with the experimental observations.
In addition, we find that also for $p_{2}/p_{1}>1$, $\chi_{\rm c}$ is decreased
even if the electrostatic screening is weak.
In order to confirm this behavior in experiments, only the outer solution
should be diluted in order to yield $p_{2}/p_{1}>1$.
Controlled experiments done on asymmetric
charged  membranes with asymmetric in and out ionic strengths are needed to confirm the predictions of the present work.
Because the lipid composition in each leaflet prepared by gentle
hydration or electroformation is not well controlled,
it is better to prepare vesicles by transferring water-in-oil droplets coated by
lipids from an oil phase to a water phase~\cite{PFW,HMKKVT} or by using asymmetric
Montal-Mueller planar bilayers~\cite{CK,MM}.

\begin{figure}[t!]
\begin{center}
\includegraphics[scale=0.6]{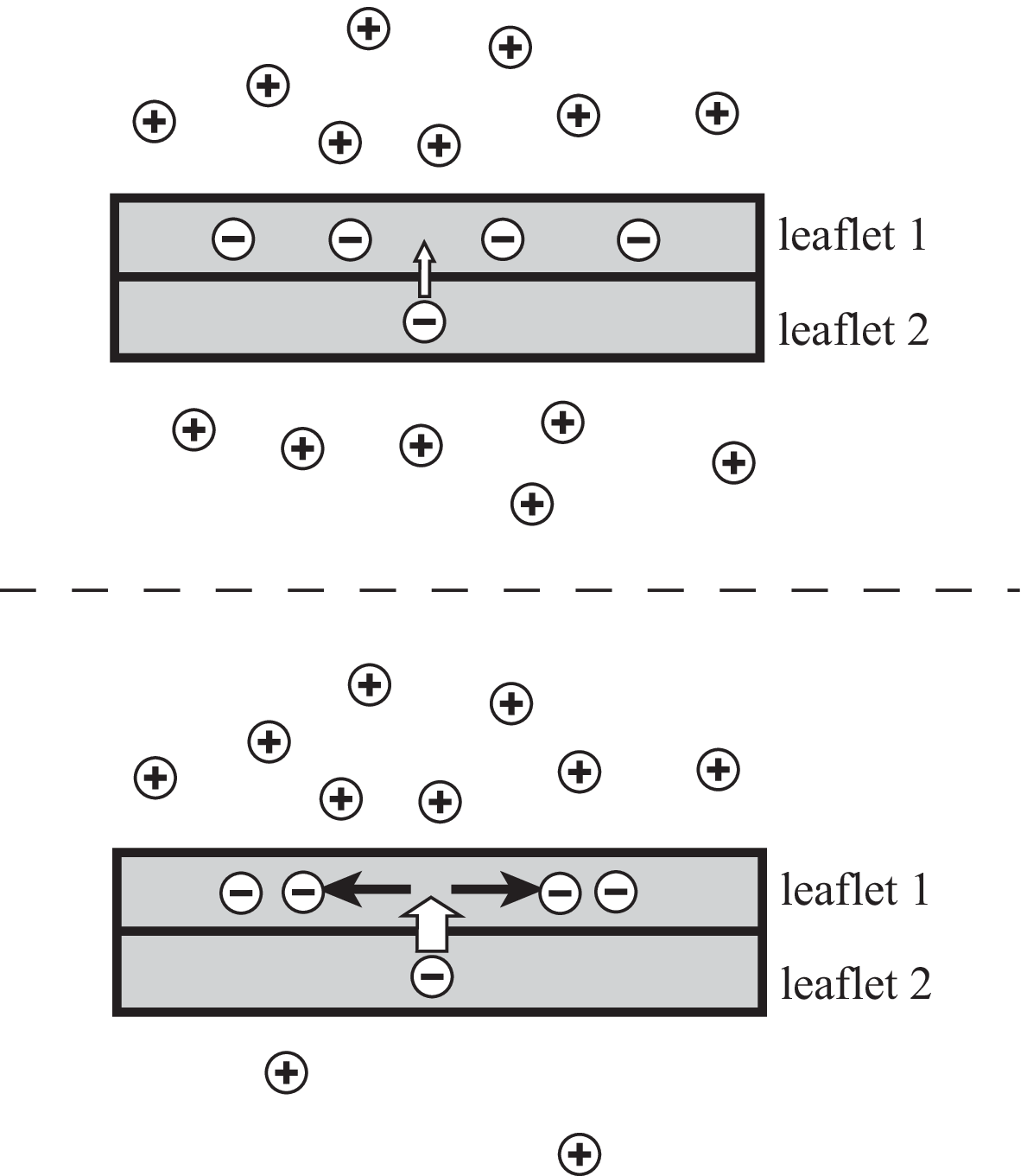}
\end{center}
\caption{\textsf{
Schematic illustration of the phase separation for an asymmetric
charged lipid bilayer when $p_{2}/p_{1}>1$.
On the upper panel, the salt concentration in solution 2 is  high (but still obeying  $p_{2}/p_{1}>1$).
On the lower panel, the salinity in solution 2 is reduced resulting in an increase of $p_2/p_1$.
The electrostatic coupling becomes then stronger
and the phase separation is induced.
}}
\label{fig6}
\end{figure}

In the following, we discuss the physical reason why $\chi_{\rm c}$ decreases for $p_{2}/p_{1}>1$.
Such a tendency is more pronounced for larger $H\sim \varepsilon_{\rm L}/d$, because the
electrostatic coupling between the two leaflets becomes stronger.
Notice that the electrostatic coupling depends on $p_{1}$ and $p_{2}$
as well as on $H$.
As schematically presented in Fig.~\ref{fig6}, the electrostatic
coupling is weaker when the salt concentration in solution 2 is
higher (upper panel of Fig.~\ref{fig6}), and
the phase separation is suppressed.

On the other hand, when the salt concentration in solution 2 is low (lower panel of
Fig.~\ref{fig6}),
the screening is not as efficient, and phase separation is induced
due to the strong electrostatic coupling.
Note that in the latter case, the strong demixing in leaflet 1 is induced by
changing the salt concentration in other reservoir (solution 2),
which is coupled directly with the opposite leaflet.
This is indeed an interesting situation because the higher salinity
in solution 1 relative to 2 triggers the phase
separation in leaflet 1 due to the strong inter-leaflet electrostatic coupling.
The region where the tie-lines are almost parallel to the $\phi_{1}$ axis means that 
the phase separation mainly takes place in leaflet 1.
As can be seen in Fig.~\ref{fig4}~(c) and (f) as well as in Fig.~\ref{fig5}~(c) and
(f), such an region becomes much larger than the region where the phase separation strongly 
occur in leaflet 2 for large $p_{2}$.
On the other hand, the phase separation also occurs  in leaflet 2,
although the concentration difference between two phases is very small.
As shown in Fig.~\ref{fig3}, $\chi_{\rm c}$ is slightly decreased when $p_{1}$ is smaller, and
can be understood by noting that the screening sufficiently weakens the electrostatic coupling.

In Figs.~\ref{fig2} and \ref{fig3} we have shown that $\chi_{\rm c}$
is smaller for larger $H$, i.e., when the electrostatic interaction
across the membrane is large.
Because of the electrostatic coupling, the local charge accumulation in
one of the two leaflets suppresses the charge accumulation in the other
leaflet, and
it is energetically unfavorable for the charged domains on each leaflet
to face each other.
As a result, a phase separation is induced in the other leaflet driven by
the electrostatic coupling.

Next we elaborate on the physical meaning of the slope of the
tie-lines in Figs.~\ref{fig4} and \ref{fig5}.
When the tie-lines are located close to the three-phase region
(triangle) and/or when $H$ is large, the tie-lines tend to be almost orthogonal to the principal diagonal.
These tie-lines connect two points on the binodal line that have a large
compositional asymmetry between $\phi_{1}$ and $\phi_{2}$. It indicates
that the charged domains in one leaflet do not prefer
to face the charged domains in the other leaflet.
The model by Baciu and May~\cite{BM} as well as the present model deal with the
compositional coupling between the two leaflets arising exclusively from
electrostatic interactions.
Consequently, it is energetically favorable for the charged domains on each
leaflet not to face each other.

In order to see the contribution of the electrostatic coupling clearly,
we consider the electrostatic coupling in the symmetric case ($p_{1}=p_{2}$).
The free energy of the inter-leaflet interaction $f_{\rm coup}$ is obtained by
subtracting the contributions of individual leaflets from the total free energy
\begin{equation}
\begin{split}
\label{coupling}
f_{\rm coup}(\phi_{1},\phi_{2})=f_{\rm tot}(\phi_{1},\phi_{2}) & -f_{\rm mix}(\phi_{1})-f_{\rm mix}(\phi_{2}) \\
 & -f_{\rm m}(\phi_{1}) -f_{\rm m}(\phi_{2}).
\end{split}
\end{equation}
For small composition differences, the potential difference across the membrane is approximately given by
\begin{equation}
\label{approx_psi}
\Delta \Psi \approx -\frac{2 p}{4Hp+\sqrt{1+p^{2} \phi_{\rm av}^{2}}}(\phi_{1}-\phi_{2}),
\end{equation}
where $\phi_{\rm av}=(\phi_{1}+\phi_{2})/2$ is the average composition
of the charged membrane.
By substituting Eq.~(\ref{approx_psi}) into Eq.~(\ref{coupling}),  the free energy of
the inter-leaflet interaction can be expanded~\cite{May} in terms of $\phi_{1}-\phi_{2}$,
\begin{equation}
\label{lambda}
f_{\rm coup}(\phi_{1},\phi_{2}) \approx -\frac{2Hp^{2}}{1+p^{2}\phi_{\rm av}^{2}+4Hp\sqrt{1+p^{2}\phi_{\rm av}^{2}}}(\phi_{1}-\phi_{2})^{2}.
\end{equation}

From Eq.~(\ref{lambda}), it is clear that $f_{\rm coup}$ is negative and
becomes smaller when the difference between $\phi_{1}$
and $\phi_{2}$ is large, because the coefficient of $(\phi_{1}-\phi_{2})^2$ is negative.
This is consistent with the diagonal tie-lines in the phase diagrams, and
a similar tendency can be seen even for the more general asymmetric systems.

%fig6

In general, the physical origin of the coupling between the two
leaflets can vary and does not have to rely exclusively
on electrostatics interaction.
Examples of such non-electrostatic effects
were considered in previous works~\cite{LA,May} and include cholesterol flip-flop, dynamic chain interdigitation,
van der Waals interaction or composition-curvature coupling.
The coupling between the two leaflets can be expressed generally by
\begin{equation}
\label{lambda1}
f_{\rm coup}=\Lambda(\phi_{1}-\phi_{2})^{2}
\end{equation}
as shown in Eq.~(\ref{lambda}).
The coupling coefficient, $\Lambda$, becomes negative in the case of the electrostatic coupling.
However,  Wagner \textit{et al.}~\cite{WLM} analyzed a model
for coupled bilayers containing such a phenomenological term as in Eq.~(\ref{lambda1})
but with $\Lambda>0$.
It is, therefore, important to consider the sign and typical value
of $\Lambda$ for different systems containing electrostatic and non-electrostatic
couplings and to reveal how the interactions
between the two leaflets contribute to the coupling constant, $\Lambda$.
Further experimental and theoretical investigations are required to gain more
insight on this coupling and its origin.

The present work is concerned with macro-phase separation in
lipid membranes.
Membranes consisting of neutral and charged lipids can  form also
modulated phases~\cite{GA} that exhibit characteristic periodic structures.
Hirose \textit{et al.}~\cite{HKA} considered the coupling between two micro-phase
separated bilayer leaflets. Various patterns, phases and their phase transitions
were predicted by assuming a phenomenological coupling between
the two leaflets.

We would like to close by mentioning again the two main approximations
used in the present model.
(i) We have treated the two leaflets as undeformable flat sheets.
By doing so we neglected any possible bilayer deformation that can occur because of
osmotic pressure
difference resulting from a difference in ionic concentrations ($p_1\ne p_2$) on
the two membrane sides.
In experiments,  phase-separated domains produce budding
toward the interior of the vesicle and this is a direct consequence of
the osmotic pressure on the
spontaneous curvature~\cite{SHSY}.
In future studies it may be of advantage
to consider the combined effect of the
phase separation and membrane deformation.
(ii) We have used the Poisson-Boltzmann theory for
symmetric monovalent salt in order to describe the electrostatic potential.
It is known that such a mean-field theory does not treat correctly
ion-ion correlation, especially when multivalent cations are involved.
This important effect should also be taken into account in follow-up modeling.

In conclusion, we have considered the phase separation in a charged
and asymmetric lipid bilayer, located between two ionic solutions having different
ionic strengths.
Considering the electrostatic effects on the phase separation, we
have introduced the electrostatic coupling between the two leaflets.
We studied the effect of the ionic strength difference between the
two solutions, $p_1\ne p_2$, and the electrostatic coupling, $H$, on the membrane phase behavior.
When the electrostatic coupling $H$ is introduced,
$\chi_{\rm c} \sim 1/T_{\rm c}$ becomes smaller than for the uncoupled case.
Moreover, even if the screening effect is small, $\chi_{\rm c}$
decreases by increasing the asymmetry in the salt concentrations.
An increase in the ionic strength leads to an enlarged coexistence
region in the phase diagram.

%%%%%%%%%%%%%%%%%%%%%%
%  Acknowledgments   %
%%%%%%%%%%%%%%%%%%%%%%

\begin{acknowledgments}

We are grateful to S. May and M. Schick for numerous comments and suggestions.
N.S. was supported by a JSPS Research Fellowship for Young Scientists
(23-8690) and Excellent Young Researchers Overseas Visit Program from
the Ministry of Education, Culture, Sports, Science and Technology
of Japan.
This work was partially supported by KAKENHI (Grant-in-Aid for Scientific
Research) on Priority Area ``Soft Matter Physics'', grant
no.\ 21540420 from the Ministry of Education, Culture, Sports,
Science and Technology of Japan, the Israel Science Foundation (ISF)
under grant no. 231/08, and the US--Israel
Binational Science Foundation (BSF) under grant no. 2006/055.
\end{acknowledgments}

\newpage

%%%%%%%%%%%%%%
%  Appendix  %
%%%%%%%%%%%%%%

\appendix*
\section{Surface potential and charging free energy of an isolated leaflet}

%%%%%%%%%%%%%%%%%%%%%
\subsection{Leaflet electrostatic potential}
%%%%%%%%%%%%%%%%%%%%%

We present the derivation of the surface potentials $\Psi_{1,2}$
[Eqs.~(\ref{surface12}) and (\ref{surface3})] and the electrostatic free energy of an
isolated leaflet [Eq.~(\ref{isolated})].
In order to find $\Psi_{1}$, we  multiply the Poisson-Boltzmann equation~(\ref{PB}) by $\psi'_{1}(z)$
and integrate it once, yielding
\begin{equation}
\int \psi'_{1} \psi''_{1} \, {\rm d} z = \kappa_{1}^{2} \int \psi'_{1} \sinh \psi_{1} \, {\rm d} z,
\end{equation}
where $\psi'_{1}={\rm d} \psi_{1} /{\rm d} z$ and $\psi''_{1}={\rm d}^{2} \psi_{1} /{\rm d} z^{2}$.
Using the appropriate boundary conditions results in the following expression:
\begin{equation}
\label{electric_field}
\psi'_{1} = -2 \kappa_{1} \sinh \left( \frac{\psi_{1}}{2} \right) \, ,
\end{equation}
and at $z=d$,
\begin{equation}
\psi'_{1}(d)=-2 \kappa_{1} \sinh \left( \frac{\Psi_{1}}{2} \right).
\end{equation}
Using Eq.~(\ref{Gauss}) and $\psi'_{\rm 1}(d)=\psi'_{\rm 2}(0)=\Delta \Psi/d$, it follows that

\begin{equation}
\sinh \left( \frac{\Psi_{1}}{2} \right) = -\frac{2\pi\ell_{\rm B}}{\kappa_{1} \Sigma} \phi_{1}
-\frac{2\pi\ell_{\rm B}}{\kappa_{1} \Sigma} \frac{\Sigma}{4\pi\varepsilon_{_{\rm W}}\ell_{\rm B}}
\frac{\varepsilon_{_{\rm L}}}{d} \Delta \Psi.
\end{equation}
From this expression, the surface potential $\Psi_{1}$ is
\begin{equation}
\Psi_{1} = -2 \sinh^{-1} \left(p_{1} \phi_{1} + p_{1} H \Delta \Psi \right),
\end{equation}
where $p_{1}=2 \pi \ell_{\rm B}/\kappa_1\Sigma$, $H=
\Sigma\varepsilon_{_{\rm L}}/4\pi\varepsilon_{_{\rm W}}\ell_{\rm B}d$ as well as other parameters were introduced
in Sec.~\ref{model}.
The surface potential on the second leaflet, $\Psi_{2}$, can be obtained in the same way.

%%%%%%%%%%%%%%%%%%%%%%%%%%%%
\subsection{Electrostatic free-energy of an isolated leaflet}
%%%%%%%%%%%%%%%%%%%%%%%%%%%%%%%

Next we calculate the electrostatic free-energy of an {\it isolated} leaflet, Eq.~(\ref{isolated}), using the charging method.
A similar derivation can be found, e.g., in Refs.~\cite{WM,Colloid}.
This energy corresponds to the energy that a leaflet feels in solution {\it without} the electrostatic coupling to the second leaflet.
When a charged leaflet  at $z=0$  is in contact with an ionic solution situated at $z>0$,
the Poisson-Boltzmann equation gives
\begin{equation}
\psi' = -2 \kappa \sinh \left( \frac{\psi}{2} \right).
\end{equation}
This equation is the same as Eq.~(\ref{electric_field}).
The nonlinear first-order differential equation can then be integrated analytically:
\begin{equation}
\psi(z)=-2 \ln \left[1+\frac{2}{\exp(\kappa z) {\coth}[\frac{1}{2}\sinh^{-1} (p\phi)]-1} \right].
\end{equation}
From this expression, the surface potential $\Psi=\psi(z=0)$ is given by
\begin{equation}
\Psi=-2 \sinh^{-1} (p \phi).
\end{equation}
Through the charging process,  the electrostatic free-energy of an isolated leaflet is obtained
by the following integral over $\Psi$ yielding an analytical expression for $f_{\rm m}$:
\begin{equation}
\begin{split}
f_{\rm m} ( \phi ) &= -\int^{\phi}_{0} \Psi(s) {\rm d} s \\
&= 2\phi \left[ \frac{1-\sqrt{(p \phi)^{2}+1}}{p \phi} + \ln \left(p \phi + \sqrt{(p \phi)^{2} +1} \right)\right].
\end{split}
\end{equation}

\newpage
%\bibliography{lipid}% Produces the bibliography via BibTeX.
%%%%%%%%%%%%%%%%%%%%%%%%%%%%%%%%%%%%%%%%%%%%%%%%%%%%%%%%%%%%%%%%%%%%%
%  References  %%%%%%%%%%%%%%%%%%%%%%%%%%%%%%%%%%%%%%%%%%%%%%%%%%%%%%
%%%%%%%%%%%%%%%%%%%%%%%%%%%%%%%%%%%%%%%%%%%%%%%%%%%%%%%%%%%%%%%%%%%%%

%%%%%%%%%%%%%%%%%%%%%%%%%%%%%%%%%%%%%%%%%%%%%%%%%%%%%%%%%%%%%%%%%%%%%
%  Tables \& Figures  %%%%%%%%%%%%%%%%%%%%%%%%%%%%%%%%%%%%%%%%%%%%%%%
%%%%%%%%%%%%%%%%%%%%%%%%%%%%%%%%%%%%%%%%%%%%%%%%%%%%%%%%%%%%%%%%%%%%%

\end{document}